\def\lessapprox{\,\raise 0.6ex\hbox{$<$}\kern -0.75em\lower 0.47ex
    \hbox{$\sim$}\,}

\def\vu{{\bf u}}

\def\vn{{\bf n}}
\def\vMA{{\bf MA}}
\def\vMB{{\bf MB}}
\def\vMC{{\bf MC}}
\def\vMD{{\bf MD}}
\documentstyle[11pt,paspconf]{article}

\begin{document}

\title{The Statistics of the large--scale Velocity Field}

\author{Francis Bernardeau}
\affil{SPhT, C.E. de Saclay, F-91191 Gif-sur-Yvette, Cedex, France}

\begin{abstract}
A lot of predictions for the statistical properties 
of the cosmic velocity field at large-scale
have been obtained recently using perturbation theory. 
In this contribution I report the outcomes of 
a set of numerical tests that aim to check these results.
Using Voronoi and Delaunay tessellations for defining the velocity 
field by interpolation between the particle velocities in numerical 
simulations, we have been able to get reliable estimates of the 
local velocity gradients.
Thus, we have been able to show that
the properties of the velocity divergence are in very good
agreement with the analytical results.
In particular we have confirmed the $\Omega$ dependence
expected for the shape of its distribution function.
\end{abstract}


\section{Introduction}

The advent of reliable redshift-independent distance estimators lead
to an enormous growth of activity in the field of measuring and 
interpreting the peculiar velocities of galaxies. 
  The velocity field can in particular be 
fruitfully investigated by means of perturbation analysis. One
important result is the, $\Omega$-dependent, velocity-density 
relationship that follows from linear theory (see e.g. Peebles 
1980). Moreover, taking advantage of the fact that
Perturbation Theory predicts that the rotational part of the 
velocity field vanishes, Bertschinger \&
Dekel (1989) developed the non-parametric 
POTENT method in which the 
local cosmological velocity field is reconstructed from the measured 
line-of-sight velocities (Bertschinger et al. 1990).

Then, via the
velocity-density relationship it is possible to 
estimate the value of $\Omega^{0.6}/b$, where it is assumed that the 
bias of the galaxies can be simply
represented by a linear bias factor $b$ 
(see the review paper of Dekel 1994 and
references therein). 

There are however other methods that have been proposed that uses
{\it intrinsic} properties of the large--scale velocity field
to estimate $\Omega$. For example, 
Nusser \& Dekel (1993) proposed to use a reconstruction method assuming 
Gaussian initial conditions to constrain $\Omega$, while Dekel 
\& Rees (1994) use voids to achieve the same goal. 
Another approach has been proposed by Bernardeau et al. (1995) and 
Bernardeau (1994a) based on the use of statistical properties
of the divergence of the locally smoothed velocity field. 
Preliminary comparisons of the analytical predictions
with numerical simulations (Bernardeau 1994b, Juszkiewicz et al. 1995, 
\L okas et al. 1995) yielded encouraging results. However, such 
a comparison is 
complicated due to the fact that the velocities 
are only known at, non-uniformly distributed, particle locations. 

Here, I report recent results obtained by Bernardeau \& van de Weygaert (1996)
and Van de Weygaert et al. (1996)
addressing  specifically the issue of the discrete nature 
of the velocity sampling. 

\section{Theoretical results}

To start with, let me remind a few analytical results
obtained from Perturbation Theory applied to the large-scale velocity
field. I consider the statistical properties of the one-point
{\it volume} averaged velocity divergence, $\theta$,
(in units of the Hubble constant) 
when the average is made with a top-hat window function.
\begin{figure}
\vspace{7.5 cm}
\special{hscale=100 vscale=100 psfile=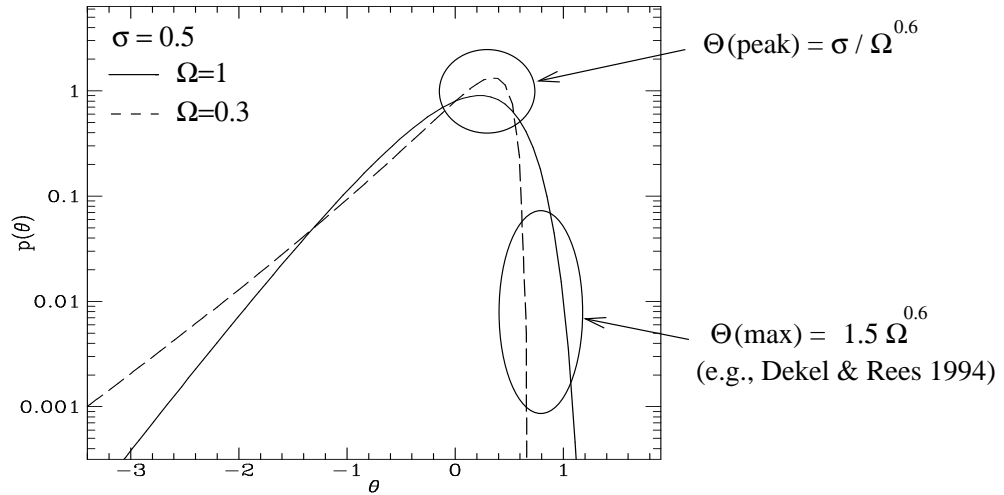}
\caption{Sketch of the PDF of the velocity divergence as given by Eq. (1)
for $\Omega=0.3$ and $\Omega=1$.} \label{fig-1}
\end{figure}
Although the first analytical results that have been obtained dealt with 
the values of high order moments of its
distribution function (Bernardeau et al. 1995, Bernardeau 1994a)
it is more convenient to consider its global shape.
Particularly interesting is the $n=-1$ case (where $n$ is the index
of the power spectrum) for which there is a simple analytical fit for 
the PDF (Bernardeau 1994b),
\begin{equation}
p(\theta){\rm d}\theta={([2\kappa-1]/\kappa^{1/2}+
[\lambda-1]/\lambda^{1/2})^{-3/2} 
\over \kappa^{3/4} (2\pi)^{1/2} \sigma_{\theta}}
\exp\left[-{\theta^2\over 2\lambda\sigma_{\theta}^2}\right]
{\rm d}\theta,
\end{equation}
with 
\begin{equation}
\kappa=1+{\theta^2\over 9\lambda\Omega^{1.2} }\,,\qquad \hbox{and} \qquad
\lambda=1-{2\theta\over 3\Omega^{0.6}}\,,
\end{equation}
where $\sigma_{\theta}$ is the variance of the distribution.
The resulting shape of the PDF is shown in Fig. 1. It is worth noting
that the $\Omega$ dependence shows up in the shape and position of the peak
(given by $\theta_{\rm peak}$ indicated on the figure) and by
the position of the large $\theta$ cutoff ($\theta_{\rm max}$ 
on the figure).
A similar property to the later one 
was also found by Dekel \& Rees (1994) using
the Zel'dovich approximation. It allows indeed to constraint $\Omega$
from the largest expanding void. In principle all these features can be 
used to constrain $\Omega$.

\section{The Delaunay and Voronoi methods}

Before trying to apply these ideas to the data 
we have extensively checked these features
in numerical simulations. 
In usual numerical analysis, the velocity field is defined with
a momentum average of the closest particles on grid points.
This method, however, yields very poor results when a subsequent
{\it volume} average is required. 
The main reason is that the two smoothing
scales, grid size and smoothing radius, cannot be very different from
each other.  The problem is in fact to define properly the velocity 
{\it field}
(the velocity at any location in space) from the velocities of a given
set of discrete and sparse points. This is what the methods proposed by
Van de Weygaert and I are designed for.

\subsection{The Voronoi method}

\begin{figure}
\vspace{7 cm}
\special{voffset=-5 hoffset=50 vscale=100 hscale=100 psfile=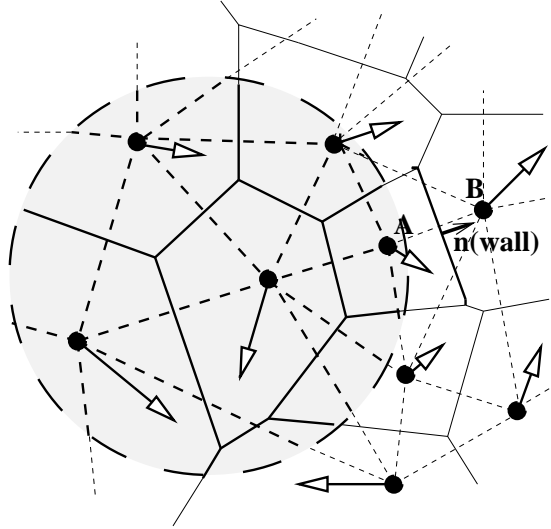}
\caption{ Voronoi and Delaunay tessellations of a 2D set of particles
(filled circles). The solid lines form the Voronoi tessellation, 
the dashed lines the Delaunay tessellation. I represented a normal vector
$\vn$ of the wall separating the points ${\bf A}$ and ${\bf B}$.} \label{fig-2}
\end{figure}

In the first method we propose the local velocity to be given
by the {\it velocity of the closest particle}. 
For defining the velocity in the 
whole space one has then to divide space in cells, each containing
a particle (of a simulation for instance), and so that
any point inside the cell is closer to it than to any
other particle. This partition is called the {\it Voronoi}
tessellation. In Fig 2. I present a 2D sketch of such a partition:
the solid lines form the Voronoi tessellation of the filled circles
representing the particles.
Then, from the initial assumption that the velocity is constant in the Voronoi
cells, the velocity gradients (in particular the divergence)
are localized on the walls. They have actually a surface density
given by ($\vu$ is the peculiar velocity)
\begin{equation}
\vu_{i,j}(\rm wall)=(\vu_{\rm ext}-\vu_{\rm int})_i\cdot\vn_j(\rm wall)
\end{equation}
where $\vn(\rm wall)$ is the unit vector normal to the wall and
going outward of the cell (see Fig. 2).
The local smoothed velocity divergences are then just
given by the sum of the fraction of all walls that are within a 
given sphere of radius $R_0$ (thick solid line in the Figure)
multiplied by the value of the divergence on each wall,
\begin{equation}
\theta_{\rm smoothed}={3\over 4\,\pi\,R_0^3}\,\sum_{\rm walls}
\,{\rm Surface}({\rm wall} \cap  {\rm sphere})\ \theta(\rm wall).
\end{equation}

\subsection{The Delaunay method}

In the Delaunay 
method the local velocity is supposed to be given by a {\it linear
combination of the velocities of the four neighbors}. If in 1D it is easy to
identify the closest neighbors, this is no longer the case in 
2D or 3D. The solution is once again provided by the Voronoi tessellation or 
rather its dual, the {\it Delaunay} tessellation. This is the triangulation
in which the particles are connected together when they share a wall 
in the Voronoi tessellation (dashed lines in Fig. 2). 
Because of the properties of the Voronoi tessellation, the 
Delaunay triangulation satisfies a criterion
of compactness: tetrahedra that are defined in such a way (or triangles in 2D)
are as less elongated as possible. This is a crucial property for doing
a subsequent linear interpolation in the tetrahedra, since
it ensures that the linear interpolation will be as good as possible.

So having identified the four neighbors $(A,B,C,D)$ of a point $M$
its velocity is assumed to be given by
\begin{equation}
\vu(M)=\alpha_A\,\vu(A)+\alpha_B\,\vu(B)+\alpha_C\,\vu(C)+\alpha_D\,\vu(D),
\end{equation}
where $\alpha_i$ are the barycentric weights of the points $(A,B,C,D)$
at the position $M$,
\begin{equation}
\alpha_A\,\vMA+\alpha_B\,\vMB+\alpha_C\,\vMC+\alpha_D\,\vMD=0,
\ \ \ \ \ \sum_{i=1}^4\alpha_i=1.
\end{equation}
The velocity gradients are then uniform in the tetrahedra 
and the local smoothed divergence is given by a sum
over all the tetrahedra that intersect a given sphere
(gray area in Fig. 2).

\subsection{Practical implementation, Validity of the methods}

In practice to use these methods we have to select only a fraction
of the points provided by the $N$-body codes. This is done
in such a way that the largest voids retain as many particles as possible.
For details see Bernardeau  and Van de Weygaert (1995). 
The tessellations are then built using the codes developed by Van de 
Weygaert (1991).

So far we have applied these methods to two different numerical simulations.
One kindly provided by H. Couchman (1991) with CDM initial condition
for $\Omega=1$ and a PM simulation with $\Omega\approx 0.3$ and a power
law spectrum of index $n=-1$ (Van de Weygaert et al. 1996)

Here I just present a very 
significant figure showing the scatter plots of the local
divergences measured in 8000 different locations (Fig. 3) with
the various available methods.
\begin{figure}
\vspace{6 cm}
\special{voffset=-10 hoffset=-20 hscale=80 vscale=80 psfile=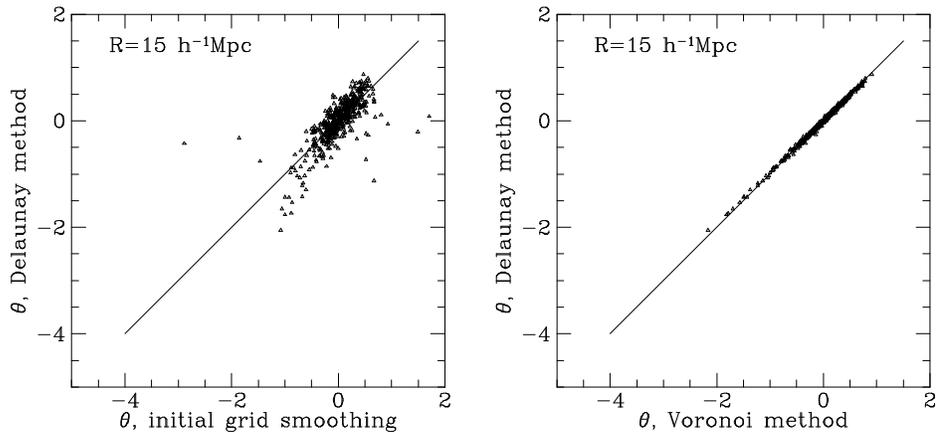}
\caption{Scatter of the local divergences measured by different methods} 
\label{fig-3}
\end{figure}
When the Delaunay method is compared to the previous grid
method the correlation is very noisy and there is even a systematic
error in the variance. When the Voronoi and the Delaunay methods are 
compared to each other, no such features are seen and there is a perfect
correlation between the two estimations.
This gives us a good confidence in our methods.

\section{The measured $\Omega$ dependence of the velocity
divergence distribution, conclusions}

\begin{figure}
\vspace{6.3 cm}
\special{voffset=-40 hoffset=-30 hscale=80 vscale=80 psfile=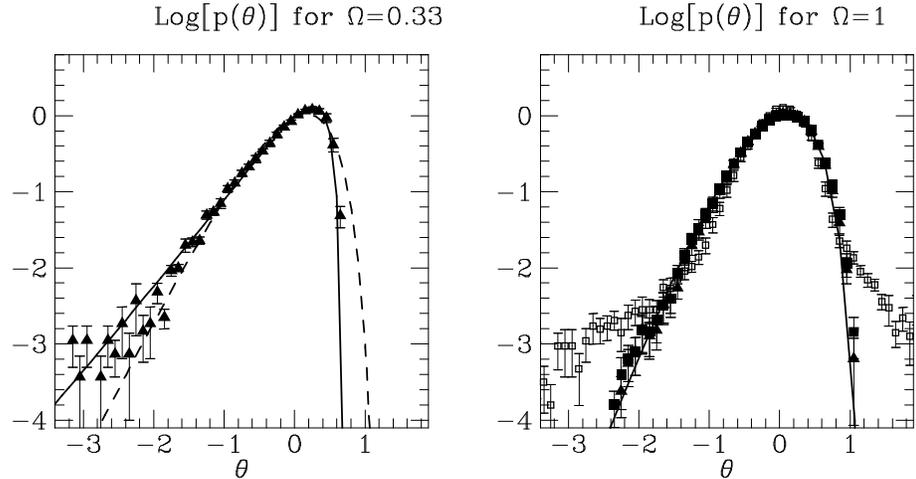}
\caption{The measured shapes of the velocity divergence PDF.
The left panel is for $\Omega=0.33$ and the right for $\Omega=1$.
The solid lines are the predictions from Eq. (1), and the dashed line for 
$\Omega=1$ with the same variance as for $\Omega=0.33$. The open squares are
from the grid method, the filled triangles and squares are 
respectively from the Delaunay and Voronoi methods.} 
\label{fig-4}
\end{figure}

The resulting PDF-s are presented in Fig. 4.
They show a remarkable agreement between the numerical estimations and the
predictions of Eq. (1). Moreover the left panel demonstrates that the
statistics of the velocity divergence is indeed sensitive to $\Omega$
(comparison of the dashed line with the solid line).

An interesting remark to make is that, in principle, it is not only
possible to  measure $\Omega$ but it must also be possible
to test the gravitational instability scenario.
The PDF given in (1) is indeed a two-parameter family of curves.
If the actual distribution fails to reproduce one of these curves
it is not compatible with the gravitational instability scenario
with Gaussian initial conditions. That would be the case, for instance, if
the observed distribution is skewed the other way around. 

However, so far, the analysis have been done essentially in numerical 
simulations. The construction of a procedure to be used with observational data
is still in progress.

\end{document}